\documentclass[11pt]{article}

\usepackage[margin=1in]{geometry}
\usepackage{multicol}

\usepackage[T1]{fontenc}
\usepackage[utf8]{inputenc}
\usepackage{microtype}

\usepackage{amsmath,amssymb,mathtools}

\usepackage{graphicx}
\graphicspath{{./}{./images/}{./figures/}}
\usepackage[percent]{overpic}

\usepackage{booktabs,tabularx,array,makecell}

\newcolumntype{Y}{>{\centering\arraybackslash}X}
\renewcommand{\arraystretch}{1.25} 

\usepackage{siunitx}
\usepackage{caption}
\usepackage{titlesec}
\titlespacing*{\section}{0pt}{12pt plus 2pt minus 2pt}{6pt plus 1pt minus 1pt}
\titlespacing*{\subsection}{0pt}{10pt plus 2pt minus 2pt}{4pt plus 1pt minus 1pt}
\usepackage[hidelinks]{hyperref}
\usepackage[capitalize,nameinlink]{cleveref}

\title{Master Oscillator and Phase Reference Line Design for the PIP-II Linac}

\author{A. Syed, B. Vaughn, P. Varghese, E. Cullerton, S. Stevenson, \\
D. Pieper, D. Peterson, J. Holzbauer, R. Madrak, A. Mosher, D. Klepec\footnotemark[1]\footnotemark[2]}

\date{October 2025}

\begin{document}
\maketitle
\begin{multicols}{2}

\footnotetext[1]{The authors grant arXiv.org and the LLRF Workshop International Organizing Committee
a non-exclusive and irrevocable license to distribute the article and certify that they have the right to grant this license.}
\footnotetext[2]{Work supported by FermiForward Discovery Group, LLC under
Contract No.~89243024CSC000002 with the U.S. Department of Energy.}

\setlength{\unitlength}{1pt}
\begin{picture}(0,0)
  \put(-72,-210){\rotatebox{90}{\small FERMILAB-CONF-25-0742-AD}}
\end{picture}

\section*{Abstract}
The phase averaging reference line system provides the RF phase reference, LO and clock signals to the LLRF and other accelerator subsystems. The PIP-II linac has RF systems at three frequencies -- 162.5\,MHz, 325\,MHz and 650\,MHz. A temperature-stabilized, low-phase-noise oscillator is used as the master oscillator. Phase reference signals at 162.5\,MHz, 325\,MHz, and 650\,MHz, along with LO signals at 182.5\,MHz, 345\,MHz, 670\,MHz and LLRF clocks at 1320\,MHz and 1300\,MHz, are generated in temperature-controlled RF modules at each frequency section. A phase reference from each module travels to the next section, where it is doubled to produce required frequencies. The reference also travels alongside the accelerating cavities in the tunnel, allowing cavity probe and reference cables to temperature track and reduce measurement errors from temperature changes or phase drift. 

\section*{1 Reference Line Operation}
In the Reference Line design, phase stabilization must address two key challenges. First, it is necessary to maintain stable relative phase between stations, where variations can arise from long cable runs and environmental factors. Second, the system must minimize absolute phase drift to ensure long-term stability of the reference signals.

\subsection*{1.1 Phase Stabilization using Averaging}
Stations along the reference line are separated by tens of meters. Thermal expansion/contraction of the transmission cables introduces delay variations that change the relative phases between stations. To reduce these effects, a forward signal is transmitted along the line and sampled at each station using directional couplers. At the end of the line, the signal is \textbf{fully reflected} by a short termination, and the reverse signal is then sampled at each station. The far-end signal drives a PLL so the feed and endpoint remain phase-matched. Averaging the forward and reverse phases improves inter-station phase stability.

\subsection*{1.2 Forward--Reflected Phase Invariance}

The forward phase at the final phase-locking coupler is anchored to the master 
oscillator reference phase, denoted $\varphi_0$, thereby locking the entire 
reference line to a stable source. For station~$n$, the forward-propagating phase 
$\varphi_{F n}$ is defined as
\begin{equation}
\varphi_{F n} = \varphi_{F1} - \beta \sum_{m=0}^{n-1} \operatorname{sgn}(m)\,l_m,
\end{equation}

\noindent
The sign function, denoted $\operatorname{sgn}(x)$, is defined as
\begin{equation}
\operatorname{sgn}(x) =
\begin{cases}
+1, & \text{if } x > 0,\\[4pt]
0,  & \text{if } x = 0,\\[4pt]
-1, & \text{if } x < 0.
\end{cases}
\end{equation}
It indicates the direction of propagation in each line segment, 
taking $+1$ for forward and $-1$ for reverse propagation.

where $\varphi_{F1}$ is the forward phase at station~1, $l_m$ is the length of 
line section~$m$ between stations $m$ and $m{+}1$, and 
$\beta = 2\pi / \lambda$ is the phase constant of the transmission line.

Similarly, the reverse (reflected) phase $\varphi_{R n}$ at station~$n$ is
{\small
\begin{equation}
\varphi_{R n} = \varphi_{F1}
  - 2\beta \sum_{m=0}^{N} \operatorname{sgn}(m)\,l_m
  + \beta \sum_{m=0}^{n-1} \operatorname{sgn}(m)\,l_m - \pi.
\end{equation}
}

where $N$ is the total number of line sections and the constant $-\pi$ accounts 
for the phase reversal introduced by the shorted end of the line.

At the phase-locking coupler, the forward phase is referenced to the master oscillator:
\begin{equation}
\varphi_{F\!pl} = \varphi_{F1} - \beta \sum_{m=0}^{N} \operatorname{sgn}(m)\,l_m = \varphi_0.
\end{equation}

Adding the forward and reverse contributions gives
\begin{equation}
\varphi_{F n} + \varphi_{R n} = 2\varphi_0 - \pi,
\end{equation}
which is independent of the individual section lengths, demonstrating phase invariance 
of the bidirectional reference line.

When the coupled forward and reverse signals, with amplitudes $A$ and $B$, 
are combined, the resulting analog-averaged output phase is
\begin{equation}
\varphi_{\text{out}} = 
\tan^{-1}\!\bigg[\frac{A \sin\varphi_{F n} + B \sin\varphi_{R n}}
{A \cos\varphi_{F n} + B \cos\varphi_{R n}}\bigg].
\end{equation}
For equal amplitudes ($A=B$), this simplifies to
\begin{equation}
\varphi_{\text{out}} = \tan^{-1}\!\big[\tan(\varphi_0)\big]
= \varphi_0 ,
\end{equation}
confirming that $\varphi_{\text{out}}$ is invariant with respect to the 
inter-station line lengths and depends only on the reference phase~$\varphi_0$.

\begin{center}
  \includegraphics[width=0.75\linewidth]{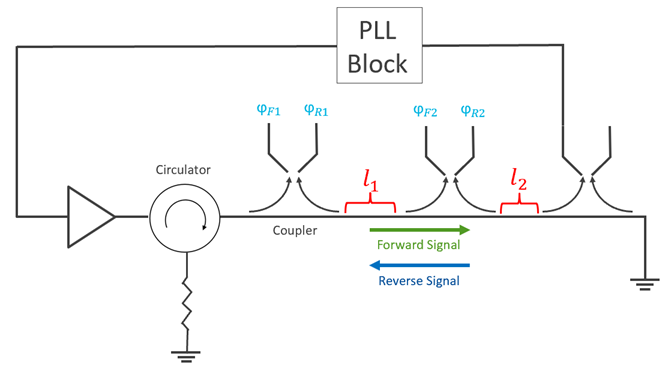}
  \captionof{figure}{Schematic of 2-station forward-reflected phase aggregated reference line.}
  \label{fig:recirc}
\end{center}

\section*{2 Design Theory}
\addcontentsline{toc}{section}{Simulation Results}

\subsubsection*{2.1 Forward--Reflected Phase Aggregation Simulation}
In the ideal case, the phase stability of the aggregated reference line is independent of inter-station line lengths. In practice, however, non-ideal return losses at component interfaces cause small reflections. After nominal transmission delays are calibrated out, these reflections appear as ripples in the output phase versus frequency, with longer line lengths producing higher ripple density and phase deviations.

To mitigate this effect, inter-station line lengths are adjusted so that the coupler output phases align with ripple peaks, where $\partial \varphi/\partial f \approx 0$. Tuning begins at the station nearest the shorted end, then proceeds upstream with iterative re-centering because adjustments interact. Lines to the right of a coupler exert stronger influence via round-trip paths, so both right and left segments are tuned until all station ripples are centered.

\begin{center}
  \includegraphics[width=1\linewidth]{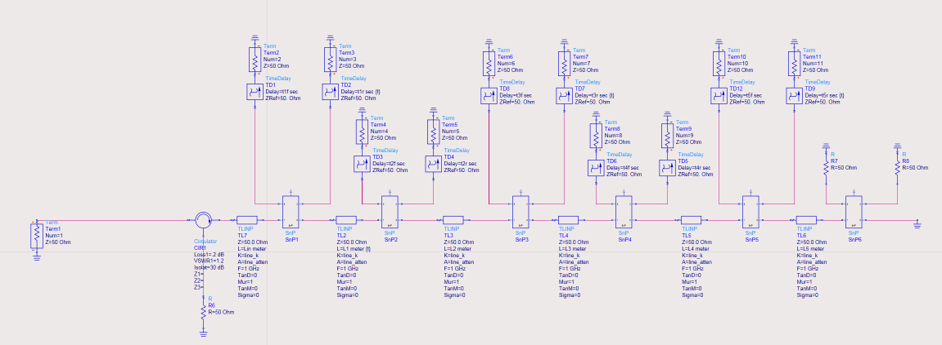}
  \captionof{figure}{ADS simulation for the reference line  }
  \label{fig:162_5MHz-sim}
\end{center}

Simulations were performed in ADS using RFS CellFlex 7/8$''$ coaxial cable models. 
The couplers were modeled using measured data: NARDA 3020A devices for the 162.5, 325, and 650\,MHz lines, 
and NARDA 3022A devices for the 1300\,MHz line. 
Referring to Figs.~2 and~3, a similar modeling and tuning procedure was applied to the 325\,MHz, 650\,MHz, 
and 1300\,MHz reference-line sections, using appropriately scaled line lengths and coupler characteristics 
to achieve equivalent forward--reflected phase aggregation and ripple alignment performance at the center frequency. 
Final line lengths are summarized in Table~1 (shown below). 
In the table, $L_n$ denotes the line immediately following station~$Lin$, 
and $L_{\text{in}}$ denotes the line from the circulator to station~1.

\vspace{-2.5 em} 
\begin{center}
\small
\captionof{table}{Optimized inter-station line lengths for each frequency section.}
\label{tab:line-lengths}
\begin{tabularx}{\linewidth}{
    >{\centering\arraybackslash}p{0.4cm}  
    >{\centering\arraybackslash}p{0.5cm}  
    >{\centering\arraybackslash}p{0.6cm}  
    >{\centering\arraybackslash}p{0.6cm}  
    >{\centering\arraybackslash}p{0.6cm}  
    >{\centering\arraybackslash}p{0.6cm}  
    >{\centering\arraybackslash}p{0.6cm}  
}
\toprule
\textbf{Freq.} & \textbf{$L_\text{in}$} & \textbf{$L_1$} & \textbf{$L_2$} & \textbf{$L_3$} & \textbf{$L_4$} & \textbf{$L_5$} \\
{\scriptsize (MHz)} & {\scriptsize (m)} & {\scriptsize (m)} & {\scriptsize (m)} & {\scriptsize (m)} & {\scriptsize (m)} & {\scriptsize (m)} \\

\midrule
\scriptsize 162.5 & \scriptsize 0      & \scriptsize 15.036  & \scriptsize 10.000  & \scriptsize 40.2585 & \scriptsize 0      & \scriptsize 0 \\
\scriptsize 325   & \scriptsize 0      & \scriptsize 15.086  & \scriptsize 15.853  & \scriptsize 15.853  & \scriptsize 15.843 & \scriptsize 15.185 \\
\scriptsize 650   & \scriptsize 15.101 & \scriptsize 32.7634 & \scriptsize 32.7634 & \scriptsize 32.7634 & \scriptsize 32.7634 & \scriptsize 15.101 \\
\scriptsize 1300  & \scriptsize 0      & \scriptsize 64.910  & \scriptsize 48.989  & \scriptsize 0       & \scriptsize 0       & \scriptsize 0 \\
\bottomrule
\end{tabularx}
\end{center}

\begin{center} \includegraphics[width=0.95\linewidth]{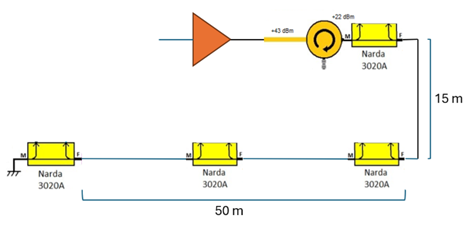} 

 \label{fig:162_5MHz_System} \end{center} \begin{center} \includegraphics[width=0.95\linewidth]{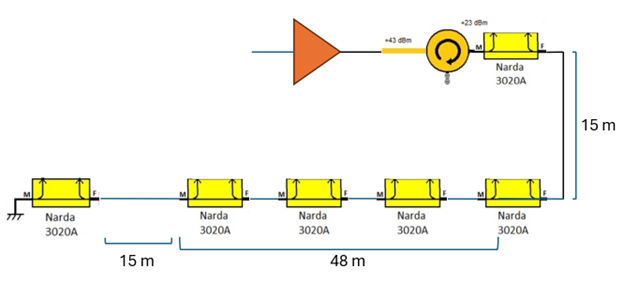} 
\label{fig:325MHz_System} \end{center} \begin{center} \includegraphics[width=0.95\linewidth]{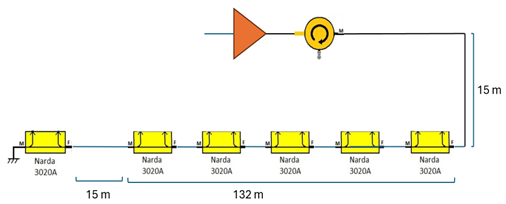} 
\captionof{figure}{Approximate lengths of the 162.5MHz, 325MHz \& 650\,MHz sections} \label{fig:650MHz_System} 
\end{center} 

\begin{figure}[htbp]
  \centering
  \begin{overpic}[width=0.8\linewidth]{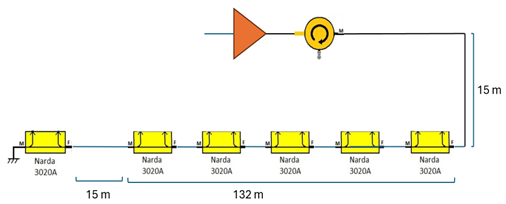}
    \put(25,85){\color{white}\bfseries 650\,MHz Section}  
  \end{overpic}
  \caption{Approximate lengths of the 650\,MHz section of the reference line.}
  \label{fig:650MHz_System}
\end{figure}

\subsubsection*{2.2 Thermal Variation Simulation}

Thermal effects in long coaxial cables impact both the \emph{physical length} of the line and the \emph{propagation constant}. 
Let each section $l_k$ denote the nominal line length at reference temperature $T_0$, with total length $L=\sum_k l_k$. 
When the temperature changes by $\Delta T=T-T_0$, the length expands with the linear coefficient of thermal expansion $\alpha$ (ppm/$^\circ$C):
\begin{equation}
l_k(T)=l_k\,(1+\alpha\,\Delta T).
\tag{8}
\end{equation}

Similarly, the phase constant $\beta$ (rad/m) varies with temperature due to dielectric velocity-factor changes, modeled by the effective temperature coefficient of delay $\gamma$ (ppm/$^\circ$C):
\begin{equation}
\beta(T)=\beta_0\,(1+\gamma\,\Delta T), \qquad 
\beta_0=\frac{2\pi}{\lambda},
\tag{9}
\end{equation}
where $\lambda$ is the guided wavelength at $T_0$.

The forward phase at station $n$, denoted $\varphi_{F n}$, is then
\begin{equation}
\varphi_{F n}(T)=\varphi_{F1}-\beta(T)\sum_{k=1}^{n-1} l_k(T),
\tag{10}
\end{equation}
where $\varphi_{F1}$ is the forward phase at the first station.  
Linearizing for small $\Delta T$ gives
\begin{equation}
\frac{d\varphi_{F n}}{dT}\approx-\beta_0(\alpha+\gamma)\sum_{k=1}^{n-1}l_k.
\tag{11}
\end{equation}

With ideal forward–reverse averaging (equal coupler amplitudes $A=B$), the temperature derivative of the combined phase is zero:
\begin{equation}
\frac{d}{dT}\bigl(\varphi_{F n}+\varphi_{R n}\bigr)=0.
\tag{12}
\end{equation}

However, if the coupler amplitudes are slightly mismatched, the imbalance can be quantified by
\begin{equation}
\varepsilon = \frac{|A-B|}{A+B} \ll 1,
\tag{13}
\end{equation}
\begin{itemize}
  
  \item $A$ and $B$ are phase matched
\end{itemize}
and under the above conditions, the residual sensitivity of the averaged output phase $\varphi_{\text{out}}$ is
\begin{equation}
\left.\frac{d\varphi_{\text{out}}}{dT}\right|_{\text{mismatch}}
\propto \varepsilon\,\beta_0(\alpha+\gamma)\,L.
\tag{14}
\end{equation}

This relation emphasizes the importance of precise amplitude matching (minimizing~$\varepsilon$), reducing the total cable length~$L$, and using low-TCD, large-diameter 7/8'' Heliax cable to achieve sub-femtosecond-level phase stability in the reference line.

\paragraph{Cable thermal data and simulation method.}
\vspace{-1.5em} 
A 7/8$^{\prime\prime}$ Heliax cable with a temperature coefficient of delay of approximately 
3~ppm/\textdegree C exhibits about 1.1~ps/\textdegree C, corresponding to roughly 
0.065\textdegree/\textdegree C of phase drift over a 100~m length at 162.5~MHz.

 To simulate thermal variation, each line section was incrementally shortened (corresponding to a reduced electrical length), and the phase difference between station~1 and each subsequent station was evaluated at every step. All coaxial cables within a line were scaled proportionally, ensuring uniform thermal contraction. The nominal phase differences relative to station~1 were then subtracted from the corresponding values at each simulated temperature step, yielding the phase deviation versus temperature relationship for all lines.

\paragraph{Simulation approach.}
\vspace{-1.05 em} 
To simulate thermal effects, each line section’s electrical length is decreased in steps corresponding to deviations from 0 to $-100$\,ppm. For each case, the relative phase difference between station~1 and each subsequent station is calculated. Because all coaxial cables are scaled by the same factor, the nominal phase differences relative to station~1 are subtracted, isolating the incremental drift due to thermal contraction. 
\vspace{-1.2em} 

\paragraph{Results.}  
\vspace{-0.8em} 
Figure~\ref{fig:thermal-all}  shows representative phase deviation plots at different frequency stations. The results confirm that ripple-centering combined with forward–reverse averaging limits drift to sub-degree levels over the expected thermal range.
\vspace{-0.7em} 

\begin{center}
  \includegraphics[width=0.8\linewidth]{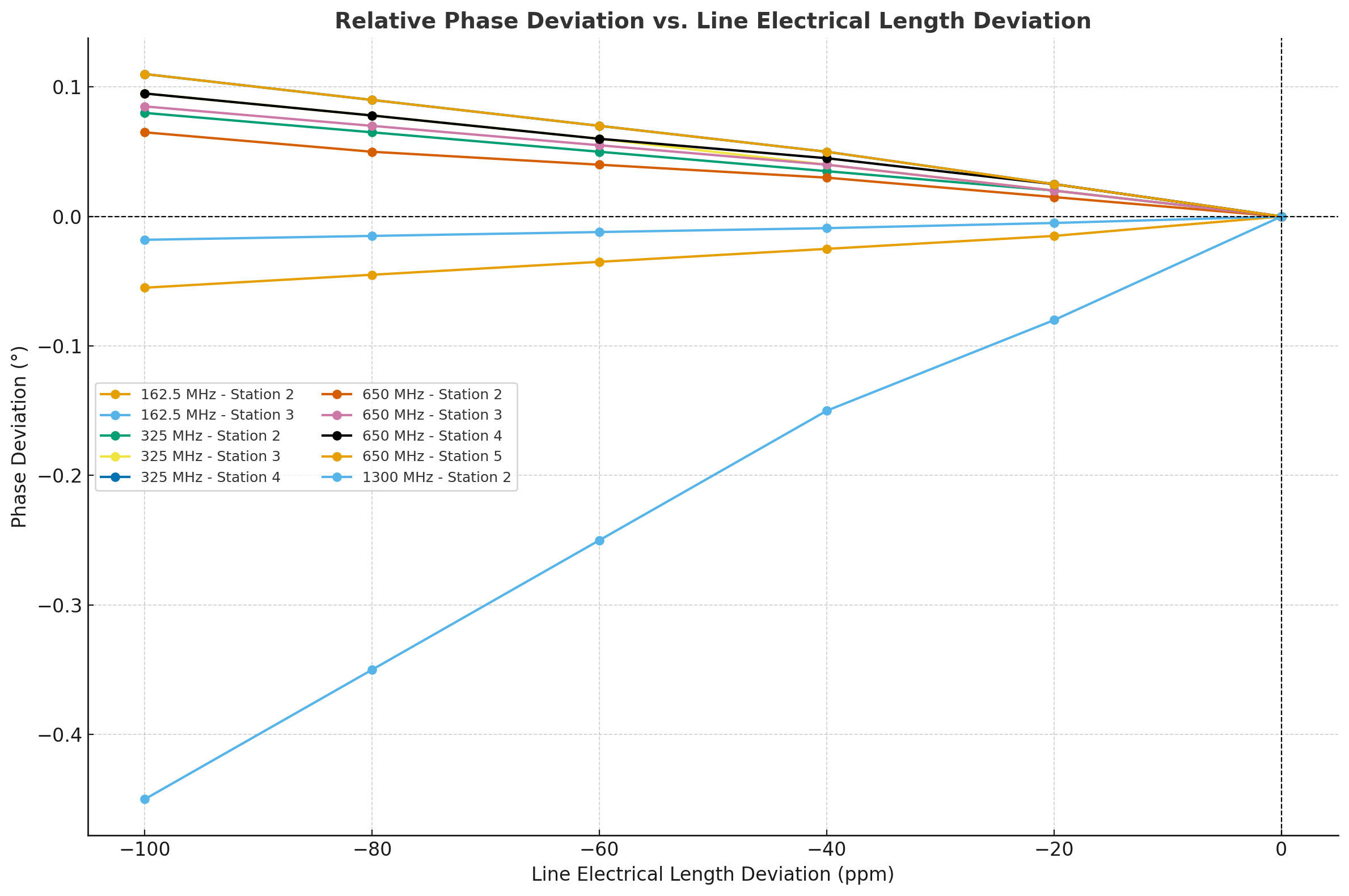}
  \captionof{figure}{Relative phase deviation with respect to station~1 as a function of electrical length change for the 162.5\,MHz, 325\,MHz, 650\,MHz, and 1300\,MHz lines.}
  \label{fig:thermal-all}
\end{center}

\section*{3 System Testing}
\addcontentsline{toc}{section}{System Testing}

\subsubsection*{3.1 Master Oscillator Measurements}

\paragraph{}
The master oscillator is the \textbf{Wenzel 10\,MHz \& 162.5\,MHz Vibration-Isolated MXO-PLMX Plate}.
Three units were procured (suffixes \textbf{A001}, \textbf{A002}, \textbf{A003}). Each provides 10\,MHz and 162.5\,MHz outputs; RF levels, spurious tones, and PLL performance were verified against factory acceptance tests and project requirements.

\paragraph{3.1.1 Phase Noise Measurements.}
\vspace{-.4em} 
Phase noise, expressed in dBc/Hz, is the single-sideband noise power density at a given frequency
offset from the carrier. The offset frequency (in Hz) indicates the separation between the carrier and
the measurement point, and integrating the phase noise spectrum yields the corresponding RMS jitter
in the time domain (reported in femtoseconds or picoseconds).

\begin{center}
  \includegraphics[width=0.85\linewidth]{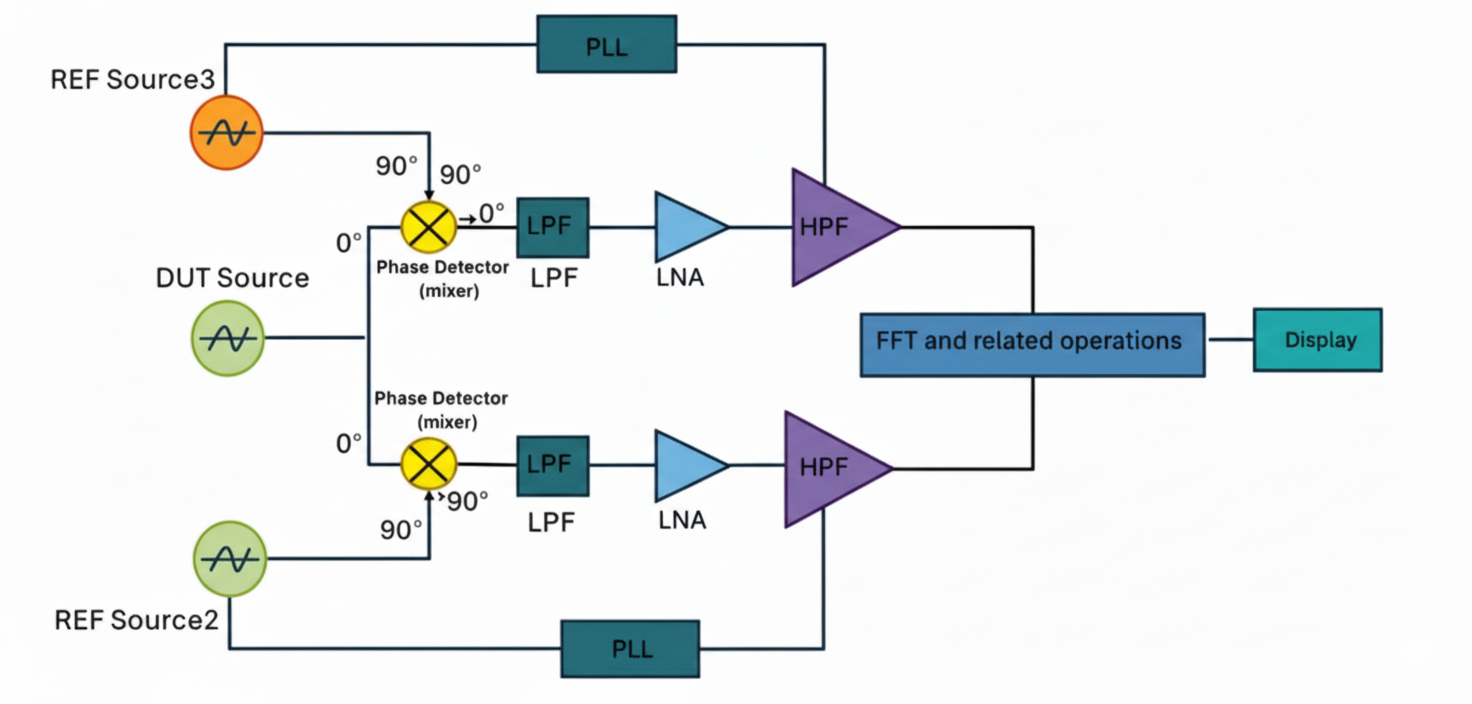}
  \captionof{figure}{Cross-correlation phase noise measurement technique.}
  \label{fig:crosscorr}
\end{center}

To characterize the oscillator units, a three-source cross-correlation method was used. 
The device under test (DUT) output was divided by a high-isolation power divider and compared 
against two independent low-noise references. Each path included a mixer and low-pass filter, 
with a phase-lock module maintaining quadrature between the DUT and the references so that 
phase variations were converted into voltage fluctuations. The baseband outputs were then 
amplified and measured simultaneously with a dual-channel FFT analyzer. 
Because incoherent mixer and amplifier noise averages down across the two channels, this 
approach improves sensitivity by approximately 15--20\,dB compared to the conventional 
two-source method.


\begin{center}
\small
\captionof{table}{Phase noise measured at Fermilab (static) for 162.5\,MHz oscillators procured for PIP-II}
\label{tab:phase-noise}
\begin{tabularx}{\linewidth}{>{\centering\arraybackslash}p{0.02cm} 
                                >{\centering\arraybackslash}p{0.6cm} 
                                >{\centering\arraybackslash}p{0.8cm} 
                                *{3}{>{\centering\arraybackslash}X}}
\toprule
\textbf{} & \textbf{Offset} & \textbf{Spec} & \textbf{A001} & \textbf{A002} & \textbf{A003} \\
 & {\scriptsize (Hz)} & {\scriptsize (dBc/Hz)} & {\scriptsize (dBc/Hz)} & {\scriptsize (dBc/Hz)} & {\scriptsize (dBc/Hz)} \\
\midrule
1 & 10    & -109 & -104 & -105 & -109 \\
2 & 100   & -128 & -137 & -133 & -128 \\
3 & 1k    & -135 & -136 & -139 & -135 \\
4 & 10k   & -175 & -165 & -172 & -175 \\
5 & 100k  & -176 & -173 & -177 & -176 \\
\bottomrule
\end{tabularx}
\end{center}

\begin{center}
\small
\setlength{\tabcolsep}{6pt}
\renewcommand{\arraystretch}{1.2}
\captionof{table}{Measured RMS time jitter at 162.5\,MHz output.}
\label{tab:jitter}
\begin{tabularx}{\linewidth}{YYYY}
\toprule
\textbf{Requirement (ps)} & \textbf{A001 (fs)} & \textbf{A002 (fs)} & \textbf{A003 (fs)} \\
\midrule
1.11 & 18.81 & 19.36 & 36.86 \\
\bottomrule
\end{tabularx}
\end{center}

\section*{4. Conclusion}
The preliminary design and testing of components for the Master Oscillator and Precision Reference Line is complete and project is under construction to meet the accelerator's demanding RF signal requirements. The core of its phase stability relies on two key techniques: \textbf{forward-reflected phase aggregation} and \textbf{active recirculation}. The former uses a shorted line and directional couplers to average forward and reflected signals, a technique validated in other operational accelerators like the Fermilab 1.3GHz NML ASTA program \cite{branlard06}. This provides a robust solution for mitigating thermal expansion in cables \cite{branlard06}. The latter, active recirculation, re-introduces the reflected signal back into the line \cite{branlard06}. Both simulations and bench testing confirm that this approach is highly effective at maintaining stable phase \cite{branlard06}.

The design also incorporates several improvements from prior experience, including the use of \textbf{large diameter 7/8'' Heliax cable} for best phase stability and \textbf{frequency multiplication} from each section to avoid phase quadrant errors associated with divider circuits \cite{pip2trs}. Components are housed in temperature-controlled enclosures set at a stable 40\textdegree C, and power supply units are kept external to minimize noise \cite{pip2trs}. Through extensive testing, including phase noise and RMS jitter measurements, the master oscillator is verified to comply with all PIP-II requirements \cite{doolittle15}. The overall system design provides a reliable, high-performance solution for RF and clock signal distribution throughout the PIP-II linac \cite{schlarb14}.

\begin{flushleft}

\end{flushleft}
\end{multicols}

\end{document}